\documentclass[12pt,preprint]{aastex}

\shorttitle{How fast could a proto-pulsar rotate?}

\shortauthors{Xu, Wang \& Qiao}

\input psfig.sty

\begin{document}

\title{How fast could a proto-pulsar rotate?}

\author{R.X. Xu, H.G. Wang \& G.J. Qiao}

\affil{School of Physics, Peking University, Beijing 100871,
   China}

\email{rxxu@bac.pku.edu.cn, wanghg@bac.pku.edu.cn, gjn@pku.edu.cn}

\begin{abstract}

According to two estimated relations between the initial period
and the dynamo-generated magnetic dipole field of pulsars, we
calculate the statistical distributions of pulsar initial periods.
It is found that proto-pulsars are very likely to have rotation
periods between 20 and 30 ms, and that most of the pulsars rotate
initially at a period $< 60$ ms.

\end{abstract}


\section{Introduction}

Pulsars provide us an unusual physical condition to increase our
knowledge of the nature. As yet, one of the essential parameters,
the initial periods $P_0$ of proto-pulsars,\footnote{
Pulsars could be neutron stars or strange stars (e.g., Xu 2001).
We discuss in this paper both possibilities of proto-pulsars being
proto-neutron stars and proto-strange stars.
}%
which may reveal precious information of dynamical supernova
process, are poorly known.
Actually there are 3 efforts to estimate $P_0$ in the literatures.
1, The period $P_0$ can be found via $T=P/[(n-1)\dot
P][1-(P_0/P)^{n-1}]$ if the age $T$ and the braking index $n\equiv
{\Omega \ddot \Omega/ \dot{ \Omega}^2}$ ($\Omega=2\pi/P$ the
angular velocity of rotation) are measured (e.g., Kaspi et al.
1994), where $P$ is the rotation period observed. This method
needs constant braking indices, which are likely to vary with time
(Xu \& Qiao 2001).
2, Monte Carlo simulation is used to study the pulsar ``current''
in the magnetic field - period diagram (e.g., Lorimer et al.
1993); the initial period of ``injected'' pulsars into the
population are adjusted so as to sustain the number of pulsars
observed at longer periods.
3, Initial periods can be inferred for pulsars that reside within
composite supernova remnants which are powered by the pulsar
spindown energy (van der Swaluw \& Wu 2001).
The derived $P_0$ via these 3 ways are 10-60 ms, $\sim 300$ ms,
and 37-82 ms (sometimes several hundred milliseconds),
respectively.
In addition, Lai et al. (2001) studied the implication of the
apparent spin-kick alignment in the Crab and Vela pulsars, and
found that the initial period should be $\la 1$ ms in both the
electromagnetic rocket model and the hydrodynamic natal model, or
$<10$ s in the asymmetric neutrino emission model, in order to
produce a high kick velocity.

In this paper, we try an alternative effort to derive the initial
period through a relation between the initial period and the
dynamo-originated magnetic field of pulsars.
The magnetic field $B$ is assumed not to change in our discussion,
since both observation and theory imply that a pulsar's $B$-field
does not decay significantly during the rotation-powered phase
(Bhattacharya et al. 1992 and, e.g., Xu \& Busse 2001). Therefore
the field strength observed for a normal pulsar represents its
initial value.
On another hand, the strong magnetic field is supposed to be
created by dynamo action during the proto-pulsar stage (Thompson
\& Duncan 1993, Xu \& Busse 2001).
We may obtain the initial period if we can find an estimated
relation between $B$ and $P_0$ during the dynamo episode.

\section{The model for calculation of pulsar initial periods}

\subsection{The model}

Dynamo action may occur during the first a few seconds of both
neutron stars and strange stars (Thompson \& Duncan 1993, Xu \&
Busse 2001).
In the magnetohydrodynamic (MHD) mean field dynamo theory, the
helical turbulence (leading to the $\alpha$ effect), differential
rotation (the $\Omega$ effect), and the turbulent diffusion (the
$\beta$ effect) are combined (e.g., Moffatt 1978, Blackman 2002).
The dynamo-created magnetic fields are complex and not very
certain for specific astrophysical processes due to the
nonlinearity of the MHD equations.
Nevertheless it is generally believed (Blackman 2002 and
references therein) that a seed field begins to grow kinematically
until a saturation (energy equipartition between field and fluid)
phase when the growth slows down due to the back action of field
on fluid in a small scale (at or below a typical length of
turbulent eddies).

The dynamo-originated field energy in this small scale is provided
by both of convection and differential rotation.
The differential rotation energy is $E_{\rm dif}\sim 10
^{46}P_0^{-2}$ erg (from eqs.(8) and (37) of Xu \& Busse 2001).
For proto-strange stars, the convection energy is $E_{\rm con}\sim
4\pi R^2 L \rho v^2\sim 10^{48}v_8^2$ erg (Xu \& Busse 2001),
where $R\sim 10^6$ cm is the stellar radius, $L\sim 10^5$ cm the
convective thickness, $\rho \sim 4\times 10^{14}$ g/cm$^3$ the
density of strange quark matter, and the convective velocity
$v=v_8\times 10^8$ cm/s could be either $\sim 10^8$ cm/s for a
large-scale convection or $\sim 10^5$ cm/s for a local turbulent
eddy.
For proto-neutron stars, a similar estimate gives $E_{\rm con}\sim
10^{48}$ erg (Thompson \& Duncan 1993).
We assume that the differential rotation energy dominates, and
thus neglects the convective energy, during the episode when
dynamo action works, since $E_{\rm dif} \ga E_{\rm con}$ as long
as $P_0\la 100$ ms.
The saturation magnetic field in small scale, $B_{\rm sat}$,
satisfies
\begin{equation}
{B_{\rm sat}^2\over 8\pi}\sim {E_{\rm dif}\over R^3}.
\end{equation}
Since $E_{\rm dif}\propto P_0^{-2}$, we come to
\begin{equation}
B_{\rm sat}=AP_0^{-1},
\label{Bsat}
\end{equation}
where $A$ is a constant for a pulsar-like compact star with
typical mass and radius.

The growth of a much large scale magnetic field (e.g., the
observed dipole one, $B_{\rm d}$) by the $\alpha$ effect only is
almost impossible because the magnetic helicity conservation
eventually leads to a so called ``catastrophic'' quenching.
Nonetheless, differential rotation may play a great role to
generate large scale fields since it can regenerate a larger scale
toroidal field.
Assuming that the dipole magnetic field, $B_{\rm d}$, is a
constant fraction of the saturation one, we have
\begin{equation}
B_{\rm d}=aP_0^{-1}.
\label{a}
\end{equation}
However, it is possible that the ratio of $B_{\rm d}$ to $B_{\rm
sat}$ is a function of $P_0$.
The Rossby number, $R_{\rm o}=P_0/\tau_{\rm con}$ (the convective
overturn time $\tau_{\rm con}\sim 1$ ms for both proto-neutron
stars and proto-strange stars), is an effective dimensionless
number to describe the importance of rotation, and may also be
representative of differential rotation.
Therefore the ratio may depend on a function of $R_{\rm o}$ since
differential rotation is essential for large sale field
generation.
Observationally, the magnetic activity, which reflects the field
strength in the atmospheres, of G-M dwarf stars correlate with
$R_{\rm o}^{-0.55\pm 0.10}$ (Simon 1990).\footnote{
The energy source for field creation in these dwarf stars could be
mainly the convective energy, which (and thus $B_{\rm sat}$) may
be less relevant to the spin period.
}
Considering eq.(\ref{Bsat}), we can thus assume
\begin{equation}
B_{\rm d}=a'P_0^{-1.5}.
\label{a'}
\end{equation}
Eqs.(\ref{a}) and (\ref{a'}) are two estimated relations between
field and initial period, which will be used respectively to
derive pulsar initial period in section 2.2.

After the proto-pulsar phase when dynamo action works, a pulsar
may spindown because of the braking torques due to magnetodipole
radiation and the unipolar generator (Xu \& Qiao 2001), with a
rate
\begin{equation}
{\dot \Omega}=-{2R^6B_{\rm d}^2\over 3c^3I}\Omega^3\eta,%
\label{dotO}%
\end{equation}
where $I\sim 10^{45}$ g cm$^2$ the momentum of inertia,
$\eta\in(1,3)$ being a function of $B_{\rm d}$, $P$, and
inclination angle $\alpha$. For the simplicity, we assume $\eta=1$
in this paper, which is also the case of conventional magnetic
dipole model for pulsars with $\alpha=90^{\rm o}$.
We then have the magnetic field $B_{12}=B_{\rm d}\times 10^{12}$ G
by
\begin{equation}
B_{12}={1\over 10^{12}}\sqrt{3Ic^3P\dot P\over 8\pi^2R^6}
=3.2\times 10^7\sqrt{P\dot P},%
\label{B}
\end{equation}
The magnetic field is assumed not to change; the field in
eq.(\ref{B}) is thus the same as in eqs.(\ref{a}) and (\ref{a'}),
by which the initial period can be obtained.

We can also have the rotation period $P(t)$ as a function of time
$t$, by integrating eq.(\ref{dotO})
\begin{equation}
P(t)^2=P(0)^2+\zeta B_{12}^2t,%
\label{Pt}%
\end{equation}
where $P(0)=P_0$, and
$$
\zeta={16\pi^2R^610^{24}\over 3c^3I}\simeq 1.95 \times
10^{-15}~{\rm s}.
$$
Time derivative of eq.(\ref{Pt}) leads to $\zeta
B_{12}^2=2P(t)\dot P(t)$, we then have the age $T$,
\begin{equation}
T={P^2-P_0^2\over 2P\dot P}.%
\label{T}%
\end{equation}
Eq.(\ref{Pt}) may be useful to study pulsar ``current'' in the
$P-\dot P$ diagram.

\subsection{Calculations based on pulsar data}

There are three approaches to obtain pulsar ages. The first is the
characteristic age $\tau=P/2\dot{P}$, which could be derived from
eq.(\ref{T}) if $P_{0}\ll P$. Second, for a pulsar associated with
a supernova remnant (SNR), the kinematic age of the SNR could be
regarded as an estimation to the pulsar age. Third, when an SNR is
identified to be the remnant of a historically recorded supernova,
the age of the pulsar within the SNR is then precisely known.

Among the youngest pulsars which locate in SNRs (see table 2 of
Camilo et al. 2002), only the Crab pulsar is confidently believed
to be associated with SN1054, the other two, PSR J1181-1926 (in
the SNR G11.2-0.3) and PSR J0205+6449 (in the SNR 3C58) are
suggested to be probably associated with SN1181 and possibly with
SN386, respectively (Clark \& Stephenson 1976). Therefore, only
the real age of the Crab pulsar is exactly known in all the
discovered pulsars.

For the Crab pulsar, with $P=33.4$ ms, $\dot{P}=4.2\times10^{-13}$
s/s and $T=948$ yr, its initial period is then $P_{0}=16.8$ ms
through eq.(\ref{T}).
Calibrated by the Crab pulsar, eq.(\ref{a}) can then be rewritten
as
\begin{equation}
B_{12}=0.0637 P_{0}^{\rm -1},
\label{B12}
\end{equation}
which provides a way to calculate the initial period.
The ATNF pulsar catalogue has released $\sim$1,300 radio pulsars
with $P$ and $\dot{P}$ measured
(http://www.atnf.csiro.au/research/pulsar/catalogue). In our
calculation, millisecond pulsars are excluded, for their magnetic
fields may decay significantly during the recycling phase. In this
paper, the criteria used to exclude a millisecond pulsar is (a)
$P\la 60$ ms, (b) $T\la 10^{9}$ years, and (c) $B\sim 10^{8-9}$ G.
Besides millisecond pulsars, those pulsars in binary systems and
those having negative values of $\dot{P}$ are also excluded.
Totally 1,069 pulsars are selected for our calculation.

The initial periods are derived via eqs.(\ref{B}) and (\ref{B12}).
The result is presented by the histograms for $P_{0}$ (left panel
of Fig.1). It is found that 60\% pulsars have initial period less
than 60 ms, exhibiting an peak at the interval 20-30 ms, as shown
in Fig.1.
It is worth noting that data of 997 pulsars are shown in the
histogram. For the other 72 pulsars, the calculated $P_{0}$ is
greater than $P$, hence results in a negative value of $T$, which
means that eq.(\ref{B12}) may not be applicable to these pulsars.

Similarly we can also find the coefficient $a'=8.24\times10^{-3}$
in eq.(\ref{a'}) by the Crab pulsar calibration. Employing this
relation, the initial periods are calculated, which is presented
in the right panels of Fig.1. There are 23 pulsars which have
negative calculated $T$, so that totally 1046 pulsars are plotted
in the histograms.
The distribution peak of $P_0$ is also between 20 and 30 ms, but
more pulsars tend to have smaller $P_0$.
Statistically, the distributions of characteristic age $\tau$ and
the age $T$ do not show significant difference. This is not
surprising since most pulsars have their initial period $P_0 \ll
P$.

\section{Conclusion and discussion}

The statistical distributions of pulsar initial periods are
inferred based on the estimated relation between the initial
periods and the magnetic dipole magnetic fields.
It is found that proto-pulsars are very likely to have rotation
periods between 20 and 30 ms, and that most of the pulsars rotate
initially at a period $< 60$ ms.
The pulsar initial periods derived in this way are comparable with
the previous estimates by the first and the third methods, but are
significantly smaller than that by the second method (section 1).

When looking at the pulsar distribution in the $P-\dot P$ diagram,
one generally concludes that the magnetic fields decays at a time
scale of $10^{6-7}$ years since the number density of pulsars with
low fields and old ages are larger than that of pulsars with high
fields and young ages.
However, calculations of ohmic decay of dipolar magnetic fields by
Sang \& Chanmugam (1987) suggest that neutron star magnetic fields
may not decay significantly during the rotation-powered phases.
The field of a strange star can also hardly decay because of the
high magnetic Reynolds number and possible color superconductivity
(e.g., Xu \& Busse 2001).
Observations indicate that the field decays only if the pulsar is
in the accretion phase (e.g., Taam \& van den Huevel 1986).
If the field does not change significantly, how can we understand
the apparent ``field decay'' in the $P-\dot P$ diagram?
We may explain this discrepancy by studying pulsar current in the
$P-\dot P$ diagram, according to eq.(\ref{Pt}) with the inclusion
of $\eta\neq 1$ in different emission models. Certainly the
braking index varies when a pulsar evolves in this way.
Further studies on this topic will be necessary and interesting.

There are a few pulsars which have initial periods $\la 5$ ms in
our calculation; for instance, 5 pulsars with $P_0<5$ ms appear in
the right histogram of Fig.1 based on eq.(\ref{a'}).
This result can not be understood in the conventional opinion of
$r-$mode instability,\footnote{
The r-mode instability may spin down a nascent strange star or
neutron star to a $P_0\sim 3-5$ ms (e.g., Andersson \& Kokkotas
2001).
} %
if it is not caused by statistical error or other factors being
inherent in our $P_0$ calculations.
However, as addressed in Xu \& Busse (2001), the $r-$mode
instability may be inhibited by differential rotation and strong
magnetic field; the problem could thus be solved.

\vspace{0.2cm} \noindent {\it Acknowledgments}:
This work is supported by National Nature Sciences Foundation of
China (10173002) and the Special Funds for Major State Basic
Research Projects of China (G2000077602).
RXX wishes to sincerely acknowledge Mrs. S.Z. Peng for her help in
preparing pulsar data two years ago.

\begin{figure}

\centerline{\psfig{figure=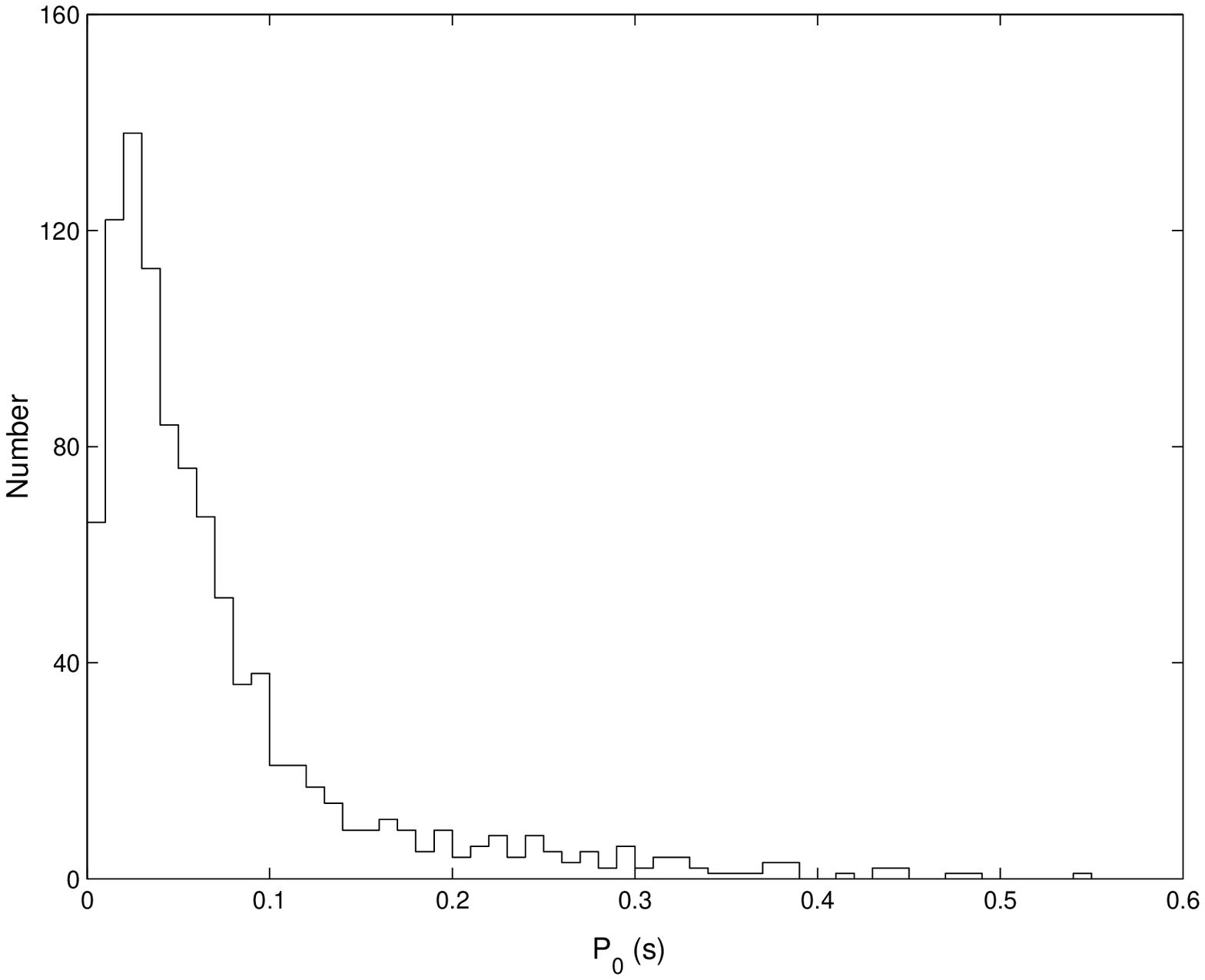,height=80mm,width=88mm,angle=0}
            \psfig{figure=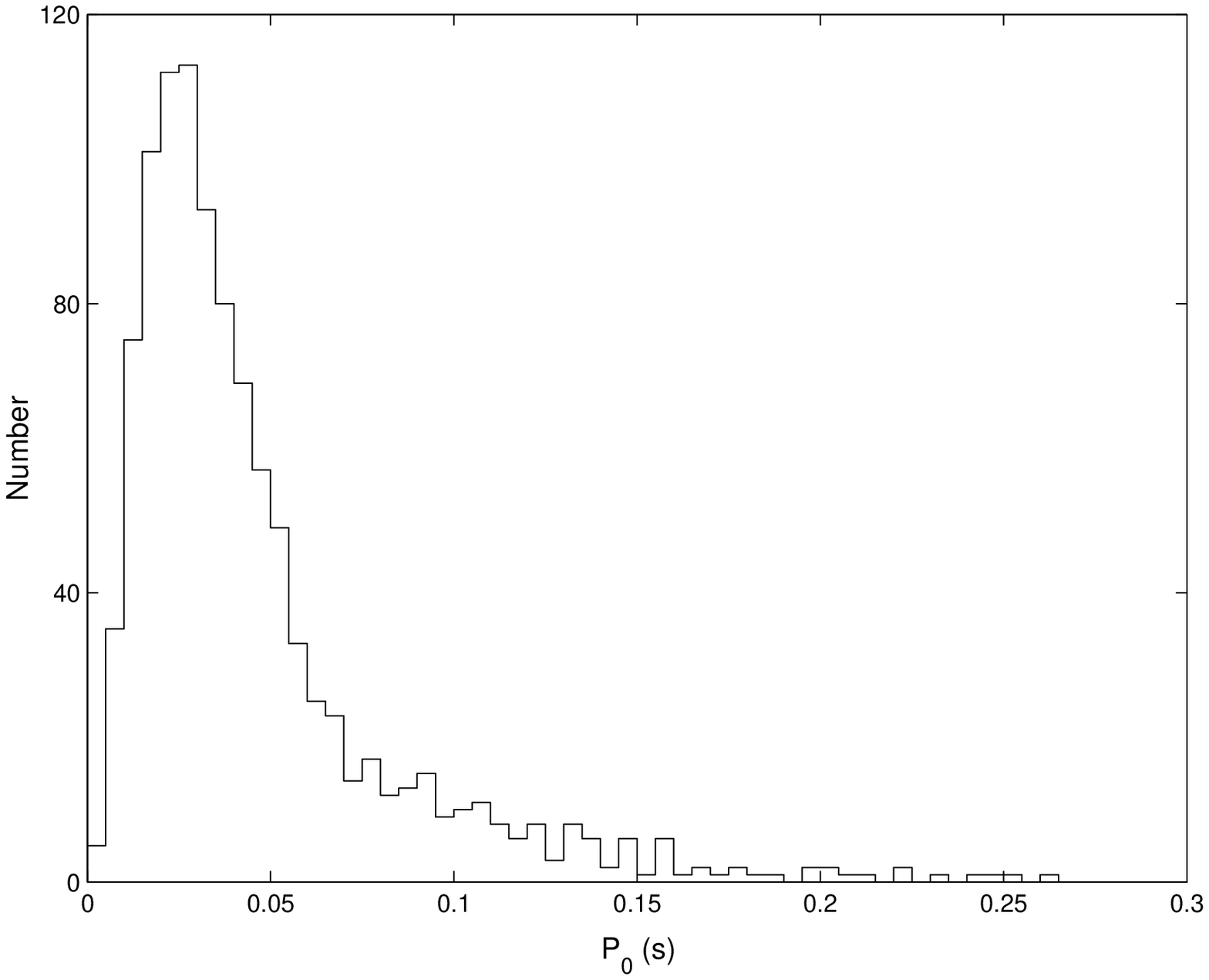,height=80mm,width=88mm,angle=0}}
\caption{Histograms for the initial period $P_{0}$. Left: $P_{0}$
is calculated from $B_{12}=0.0637 P_{0}^{\rm -1}$, in which data
of 997 pulsars are used and each bin represents an interval of 10
ms. Right: $P_{0}$ is calculated from
$B_{12}=8.24\times10^{-1.5}P_{0}^{\rm -1.5}$, in which data of
1046 pulsars are plotted and each bin represents 5 ms.
\label{Fig1}}
\end{figure}

\end{document}